\documentclass[nopubldata]{lpor2014}
\usepackage{blindtext}
\usepackage{hyperref}
\usepackage{color,soul}
\hypersetup{%
  colorlinks,
  plainpages=false,
  pdfpagelabels,
  breaklinks,
  pdfview=Fit,
  bookmarksopen,
  bookmarksnumbered,
  linkcolor=black,
  anchorcolor=black,
  citecolor=black,
  filecolor=black,
  urlcolor=LPRred
  }%
\graphicspath{{./figs/}}
\setpages[]{}
\setvolume[0]{0}
\setyear{2011}%
\setdoi{201100000}%
\oddheadlogotext{}{Review Paper}
\usepackage[Symbol]{upgreek}
\usepackage{amsmath}
\usepackage{amssymb}
\usepackage{mathrsfs}
\usepackage{graphicx}
\usepackage{dcolumn}
\usepackage{bm}
\newcommand{\rd}[1]{\textcolor{black}{ #1}}
\begin{document}

\titlefigure{toroidal}
\title{Nonseparable states of light: From quantum to classical}
\titlerunning{Nonseparable states of light}

\author{Yijie Shen\inst{1,*} and Carmelo Rosales-Guzm\'an\inst{2,3,*}}%
\authorrunning{Y. Shen and C. Rosales-Guzm\'an}
\mail{\email{y.shen@soton.ac.uk (Y.S); carmelorosalesg@cio.mx (C.R-G);}}

\institute{%
Optoelectronics Research Centre, University of Southampton, Southampton SO17 1BJ, United Kingdom
\and
Centro de Investigaciones en \'Optica, A. C., Loma del Bosque 115, Col. Lomas del Campestre, 37150, Le\'on, Gto., M\'exico
\and
Wang Da-Heng Collaborative Innovation Center for Quantum Manipulation and Control, Harbin University of Science and Technology, Harbin 150080, China}


\abstract{%
	Controlling the various degrees-of-freedom (DoFs) of structured light at both quantum and classical levels is of paramount importance in optics. It is a conventional paradigm to treat diverse DoFs separately in light shaping. While, the more general case of nonseparable states of light, in which two or more DoFs are coupled in a nonseparable way, has become topical recently. Importantly, classical nonseparable states of light are mathematically analogue to quantum entangled states. Such similarity has hatched attractive studies in structured light, e.g. the spin-orbit coupling in vector beams. However, nonseparable classical states of light are still treated in a fragmented fashion, while its forms are not limited by vector beams and its potential is certainly not fully exploited. For instance, exotic space-time coupled pulses nontrivial light shaping towards ultrafast time scales, and ray-wave geometric beams provide new dimensions in optical manipulations. Here we provide a bird's eye view on the rapidly growing but incoherent body of work on general nonseparable states involving various DoFs of light and introduce a unified framework for their classification and tailoring, providing a perspective on new opportunities for both fundamental science and applications.}

\maketitle

\section{Introduction}
\label{sec:intro}
Modern optics can be divided in several branches but all can be classified as either quantum or classical optics, featuring salient fundamental divides ~\cite{barnett2017journeys,lakshminarayanan2018mathematical}. For instance, a classical state can be measured with full certainty without being affected, at the quantum level, however, a measurement of a quantum state introduces uncertainty as it, counterintuitively, changes its properties. What is the origin of such counterintuitive phenomena? The main clue comes from \textit{nonlocality} and \textit{nonseparability} in \textit{quantum entanglement}~\cite{paneru2020entanglement,ismael2016quantum}. Nonlocality usually refers to the interaction among particles, acting at a distance, popularized by Einstein's description as ``spooky action at a distance''. Nonseparability means the state of any particle in a system cannot be separated. The nonlocality and nonseparability of entanglement seem counterintuitive but have been proved by rigorous experimental verification ~\cite{salart2008testing,hensen2015loophole,yin2017satellite}. \rd{In quantum optics, entangled systems of multiple photons with different degrees of freedom (DoFs) have been demonstrated, where the measurement of a DoF on one photon affects the states of others~\cite{horodecki2009quantum,pan2012multiphoton,wang201818}. 
In classical optics, every effect is exact locally realistic without nonlocality, but the nonseparability can be constructed by local DoFs of light to emulate quantum-like nonseparable states.}


\begin{figure*}
\centering
\includegraphics[width=0.85\linewidth]{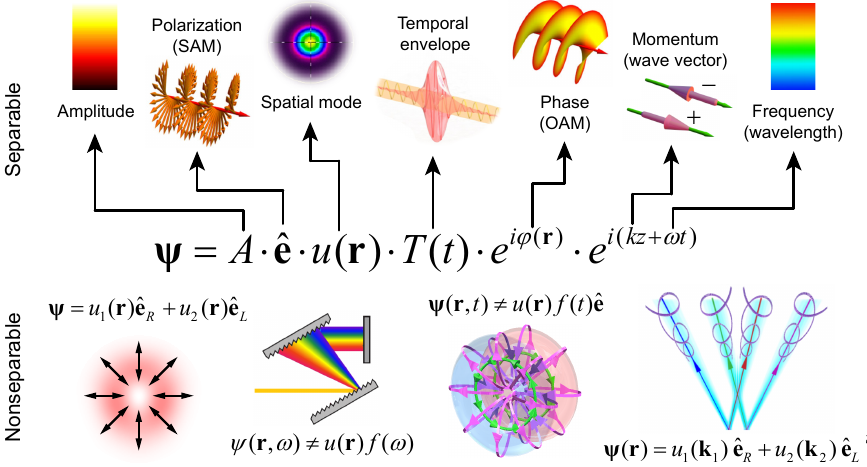}
\caption{\label{f1}{\bf\textsf{\mbox{Separable and nonseparable states of light.}}} The fundamental parameters as DoFs to control in structured light include: amplitude, polarizaiton, spatial mode, temporal envelope, phase, wave vector, and frequency. While the paradigm is to shape light structure with separable control of various DoFs. The more general light control is in its nonseparable states, such as the vector beams with nonseparability of spatial mode and polarization, space-time nonseparable structured pulse, and 
momentum-polarization nonseparable ray-wave structured light.}
\end{figure*}

Importantly, nonseparability, can also be achieved between the multiple intrinsic DoFs of a single photon, which is known as local entanglement~\cite{simon2010nonquantum,eberly2015shimony,eberly2016quantum,qian2018entanglement}. This is precisely the type of entanglement that can be used to describe \rd{the nonseparability of classical light}, \rd{that is controversially termed ``classical entanglement''~\cite{ghose2014entanglement,spreeuw1998classical,forbes2019quantum,karimi2015classical,aiello2015quantum,forbes2019classically}}. Typical examples are \rd{spin-orbit vector beams, nonseparable in the spatial and polarization DoFs~\cite{pereira2014quantum}. This existing parallelism between classical and quantum nonseparable states enable the applicability of many of the quantum methodologies to characterize some of the properties of quantum-like structured light fields, for instance, the topological phase and Bell-like inequality for spin-orbit laser modes~\cite{souza2007topological,borges2010bell}.} Hereafter we will refer to such nonseparable classical light fields as \rd{entanglement-like states and define them} as the nonseparable sum of product states of more than two DoFs of light, which exist in different vector spaces. Such definition has been widely endorsed and achieved with different DoFs in spatial, temporal, and frequency domains~\cite{simon2010nonquantum,eberly2015shimony,eberly2016quantum,qian2018entanglement,ghose2014entanglement,spreeuw1998classical,forbes2019quantum,karimi2015classical,aiello2015quantum,forbes2019classically}. Akin to quantum entangled states, such nonseparable states of any two or more DoFs gives rise to a Hilbert space that is the tensor product of the spaces corresponding to the individual DoFs. Hence, although the nonlocal action does not exist in classical optics, nonseparability is a common mathematical relationship to both \rd{quantum entangled} and \rd{classical nonseparable states of light. Based on this similarity, the methodology of quantum optics can be nicely adapted to characterize some of the properties of quantum-like structured light and is also useful in many applications~\cite{qian2015shifting,konrad2019quantum,khrennikov2020quantum,ndagano2017characterizing}.}

In mathematics, a function is nonseparable if it cannot be factorized as a product of two or more independent terms. For example, the function $F(x,y)=af_1(x)g_1(y)+bf_1(x)g_2(y)=[f_1(x)] [a g_1(y)+ b g_2(y)]$ is separable, conversely, the function $G(x,y)=a f_1(x)g_1(y)+b f_2(x)g_2(y)$ is non separable, here $a$ and $b$ are constant terms, $f_1$ and $f_2$ ($g_1$ and $g_2$) are orthogonal functions. \rd{Based on this definition, a generalized nonseparable state with $m$ DoFs can be written using Dirac's notation as,}
\begin{align}
\nonumber
\left| \psi  \right\rangle = & \sum_{i=1}^n\alpha_i\bigotimes_{j=1}^m\left|d_j^{(i)}\right\rangle\\ \nonumber
= &{{\alpha }_{1}}\left| d_{1}^{(1)} \right\rangle \left| d_{2}^{(1)} \right\rangle  \cdots  \left| d_{m}^{(1)} \right\rangle +{{\alpha }_{2}}\left| d_{1}^{(2)} \right\rangle  \left| d_{2}^{(2)} \right\rangle \cdots  \left| d_{m}^{(2)} \right\rangle  \\ 
 & +\cdots +{{\alpha }_{n}}\left| d_{1}^{(n)} \right\rangle \left| d_{2}^{(n)} \right\rangle\cdots \left| d_{m}^{(n)} \right\rangle,   \label{e1}
\end{align}
where the coefficients $\alpha_i$ ($i=1,2,\cdots,n$) are complex nonzero values that obey the relation $\sum_{i=1}^{n}\alpha_i\alpha_i^\dagger=1$, the ket $\left|d_j^{(i)}\right\rangle$ ($j=1,2,\cdots,m$) denotes the $j$-th DoF, and, importantly, any DoF substate $\left|d_j^{(i)}\right\rangle$ cannot be factorized (the values $d_j^{(i)}$ for $i=1,2,\cdots,n$ cannot be all the same). For the sake of clarity, the following definitions \rd{from quantum mechanics are given:} 

\rd{\textbf{Definition 1}: \textit{Dimensionality} is defined by the number of linearly independent eigenstates of a DoF. For example, the polarization state of light is 2-dimensional, because it has two eigenstates, that can be horizontal or vertical polarization, noted as $|H\rangle$ and $|V\rangle$. While, orbital angular momentum (OAM) state of light, $|\ell\rangle$ ($\ell=0,\pm1,\pm2,\cdots$), theoretically allow infinite dimensionality, where the eigenvalues can be arbitrary integers.}


\rd{\textbf{Definition 2}: \textit{Multi-partite entanglement} refers to the entanglement achieved between multiple nonseparable systems in a same DoF. For example, the state given by two photons entangled in the polarization DoF described by $|\psi\rangle=\frac{1}{\sqrt{2}}(|H\rangle_A|H\rangle_B+|V\rangle_A|V\rangle_B)$, is a bipartite entangled state.}


\rd{\textbf{Definition 3}: \textit{Hyper-entanglement} refers to the entanglement achieved between multiple nonseparable systems in multiple DoFs, e.g., the polarization-OAM hyper-entangled photon pair, $|\psi\rangle=\frac{1}{{2}}(|H\rangle_A|H\rangle_B+|V\rangle_A|V\rangle_B)(|\ell\rangle_A|\ell\rangle_B+|-\ell\rangle_A|-\ell\rangle_B)$~\cite{vallone2009hyperentanglement}. Further, if each system has an independent DoF, then it is called \textit{hybrid-entanglement}, e.g., the polarization-OAM entangled state, $|\psi\rangle=\frac{1}{\sqrt{2}}(|H\rangle_A|\ell\rangle_B+|V\rangle_A|-\ell\rangle_B)$~\cite{cozzolino2019air}.}

\rd{Quantum entangled states are usually constructed by different subsystems nonseparable in one or more DoFs and classified according to the number of subsystems or the number of DoFs. However, in classical domain without nonlocality, it is impossible to have nonseparability between different systems, the kind of nonseparability available is local and happens between different DoFs of the same system, as expressed by Eq.~(\ref{e1}).}

\begin{figure*}[ht]
\centering
\includegraphics[width=0.82\linewidth]{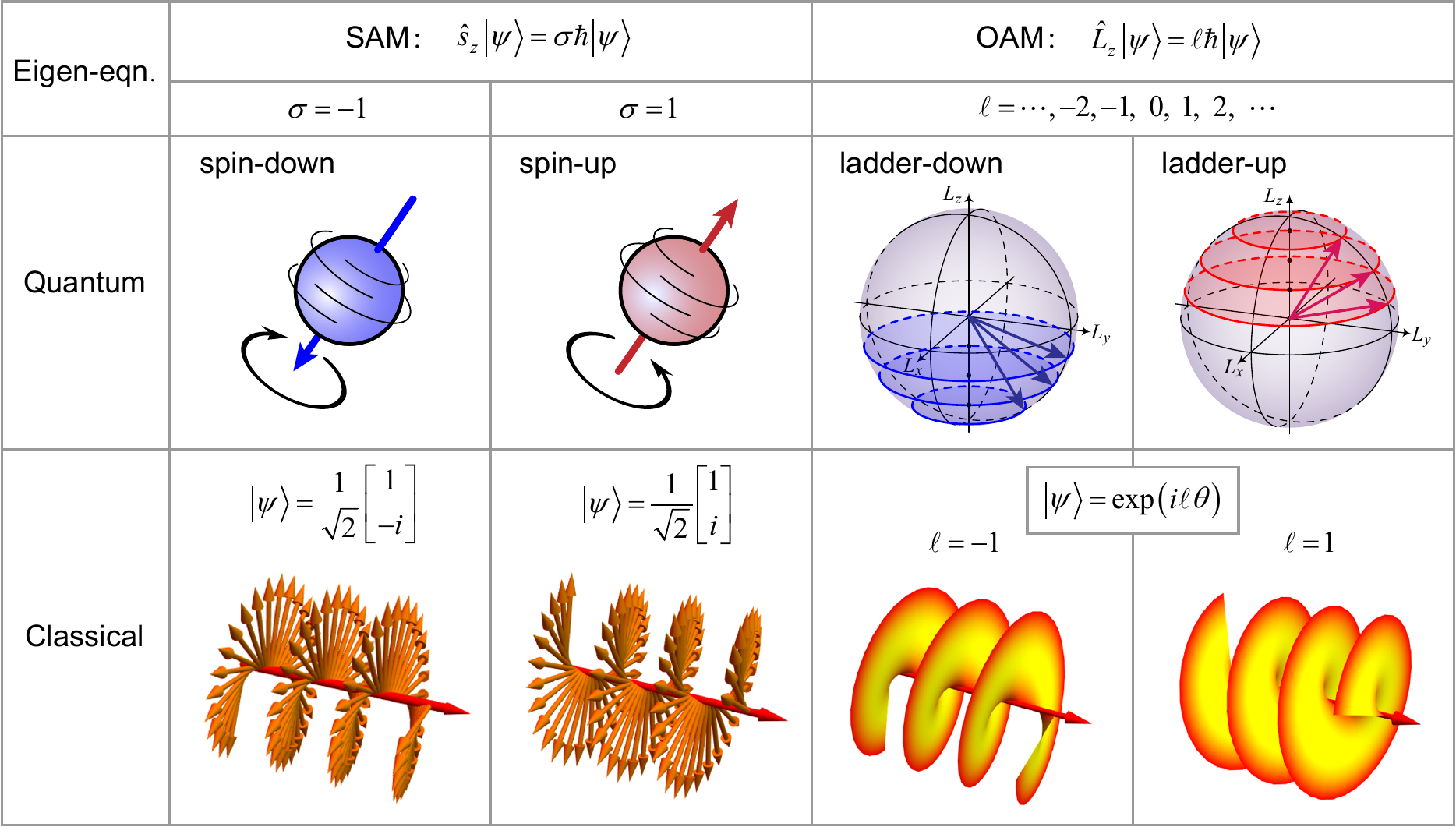}
\caption{\label{f2}{\bf\textsf{\mbox{Typical quantum-analogue models of classical light.}}} The spin operator has two eigenvalues ($\sigma = \pm 1$) with spin-up and -down eigenstates, the vector representations of which are the same as the Jones vectors of left- and right-handed circular polarizations; the OAM operator has possible infinite eigenvalues ($\ell = 0, \pm 1, \pm 2,\cdots$) with eigenstates of Hilbert factor $\exp(i\ell\theta$), which represents the vortex waves with helical isophase surfaces.}
\end{figure*}

Notably, light has various intrinsic parameters or DoF ~\cite{forbes2021structured}, such as, amplitude, polarization, spatial mode, temporal envelope, phase, wave vector, and frequency, see Fig.~\ref{f1}. A certain amount of parameters can be controlled as multiple DoFs when they can be treated independently in a optical system. However, the paradigm of light shaping is to modulate various DoFs separately. The more general light control is in its nonseparable states. For example, vector beams are the solutions of optical modes with nonseparability in their spatial mode and polarization DoF~\cite{rosales2018review}. The existence of space-time nonseparable solution of localized waves, that cannot be expressed as products of spatial and temporal functions, has been also proved~\cite{hernandez2008localized}. The spectral vector beams are examples of wavelength-polarization nonseparable states of light~\cite{kopf2021spectral}.

\rd{Here we review the rapidly growing topic of nonseparable states of classical light and some of its analogies to quantum entangled states. Although quantum entanglement is fundamentally different from classical light, nonseparability is a common property to both. Based on this analogy,} it makes sense to adapt the tools that have been already developed in quantum mechanics to analyze \rd{quantum-like nonseparable states of classical light} and demonstrate novel \rd{classical structures} that can mimic certain properties of quantum entanglement \cite{toninelli2019concepts,konrad2019quantum}. Based on a quantum-inspired methodology, we provide a unified framework for the classification of nonseparable structured light, guiding the new forms of nonseparable structured light to be controlled and providing new opportunities for both fundamental science and applications.

\section{Analogies between quantum and classical states}

In this section we revisit some of the mathematical and physical similarities between quantum and classical optics as the foundation to extend these to other complex nonseparable states. To start with, it is worth mentioning that classical analogies have been proposed for a wide range of quantum states~\cite{dragoman2013quantum}. As a first example, the quantum spin angular momentum (SAM) state corresponds to the circular polarization state of light~\cite{bliokh2015quantum}. Thus, an arbitrary polarization state can be expressed as a superposition of the two eigenstates. {As a second example, the quantum state associated to the orbital angular momentum (OAM) also has its classical analog, the vortex beam carrying a helical phase~\cite{shen2019optical} (Fig.~\ref{f2}) and a well defined amount of OAM per photon. Notably, any arbitrary phase distribution of light can be decomposed into various OAM eigenstates.}


Besides the above mentioned classical-quantum \rd{analogies} between polarization and SAM as well as vortex beam and OAM, there exist an even larger generalization of classical optical modes resembling quantum states. Actually, the basic equations for quantum states and classical modes usually share a same mathematical formation (See Supporting Information for more detailed derivations of the formal analogies). For instance, the light modes propagating in free space satisfy the basic equation, i.e. paraxial wave equation, as a form of Schr\"odinger equation just with $z$ variable replacing the time~\cite{lax1975maxwell}. 
\begin{equation}
\nabla^2_\mathbf{r}u(\mathbf{r},z)=-2ik\frac{\partial u(\mathbf{r},z)}{\partial z},\label{pwe}
\end{equation}
where $\nabla^2_\mathbf{r}=\partial^2/\partial x^2+\partial^2/\partial y^2$, $(\mathbf{r},z)=(x,y,z)$, the mode is assumed to be propagating along $z$-axis, and $u(\mathbf{r},z)$ represents the envelope function related to the electric field by $\mathbf{E}(\mathbf{r},z,t)=\hat{\mathbf{e}}u(\mathbf{r},z)e^{-i\omega t}e^{-ikz}$, where $\hat{\mathbf{e}}$ is polarization vector and $k=\omega/c$ is wavenumber. Therefore, the mathematical methods in quantum mechanics to solve Schr\"odinger equation can be transferred to solve classical mode with a proper postulate.

\textbf{Postulate}~\cite{konrad2019quantum}: A transverse mode $u(x,y)$ of paraxial light is represented by a vector $|u\rangle$ in an infinite-dimensional Hilbert space isomorphic to $\mathcal{L}^2(\mathbb{R}^2)$, the square modulus of which is associated to its transverse intensity $|u|^2=\langle u|u\rangle\propto||\mathbf{E}||^2$. In contrast, for a quantum state it represents its probability density.


According to the quantum mechanics operator algebra, we can define the position and momentum operators, $\hat{x}=x$, $\hat{y}=y$ and $\hat{p}_\mathbf{r}=-i\nabla_\mathbf{r}$, respectively,  to transform Eq.~(\ref{pwe}) in the equivalent form~\cite{nienhuis1993paraxial}, 
\begin{equation}
\frac{\text{d}}{\text{d}z}|u(z)\rangle=-\frac{i}{2k}\hat{p}^2|u(0)\rangle \Rightarrow|u(z)\rangle=\hat{U}(z)|u(0)\rangle,
\end{equation}
where the propagation operator $\hat{U}(z)=\exp{(-\frac{i}{2k}\hat{p}^2z)}$ describes the transverse mode propagation along $z$-axis~\cite{stoler1981operator} and corresponds to the time-dependent evolution of a particle state in quantum mechanics. Using such operators we can define the ladder operators as~\cite{nienhuis1993paraxial}:
\begin{equation}
\hat{a}_x=\frac{1}{\sqrt{2k}}(k\hat{x}+i\hat{p}_x),\quad \hat{a}_y^\dagger=\frac{1}{\sqrt{2k}}(k\hat{y}+i\hat{p}_y),
\end{equation}
which are of the same form of the ladder operator for Fock states or number states of photons~\cite{gerry2005introductory}. From such operators, high-order transverse modes $|u_{n,m}\rangle$ can be obtained, by applying the raising operator to the fundamental Gaussian mode $|u_{0,0}\rangle$~\cite{nienhuis1993paraxial},
\begin{equation}
|u_{n,m}\rangle=\frac{(\hat{a}_x^\dagger)^n}{\sqrt{n!}}\frac{(\hat{a}_y^\dagger)^m}{\sqrt{m!}}|u_{0,0}\rangle.
\end{equation}
The eigenfuncation of this result refers to a Hermite--Gaussian (HG) modes, the conventional laser mode solved in Cartesian coordinate system. Furthermore, the Laguerre--Gaussian (LG) modes in circular coordinate and general Hermite--Laguerre--Gaussian (HLG) modes can be easily obtained by applying additional SU(2) transformation to the ladder operators~\cite{chen2018originating,shen20202,shen2021rays}.

In addition, recent advances in the field of structured light have evinced novel kinds of quantum-analogue classical states~\cite{dragoman2013quantum,aiello2015quantum,forbes2019classically}. As a result, quantum analogies are not limited to free-space light beams. For instance, the Schr\"odinger equation and related processing methods were also extended to study the classical light behavior in fibers~\cite{shen2017gain,eslami2018new}, nonlinear media~\cite{chavez2007quantumlike,shen2018beam}, complex waveguide systems~\cite{feng2017non,ozdemir2019parity}. More typical examples and details will be reviewed in the following sections. Therefore, it is still highly appealing to explore more intriguing quantum-analogue formation in various kinds of light fields.
\section{\rd{Two-DoF nonseparable states}}
Hereinafter, we introduce a unified framework for a comprehensive classification of nonseparable states of light. We start from the classical \rd{nonseparable states with two DoFs, such as space-polarization or space-time and end with a generalized description.}

\begin{figure*}
\centering
\includegraphics[width=0.95\linewidth]{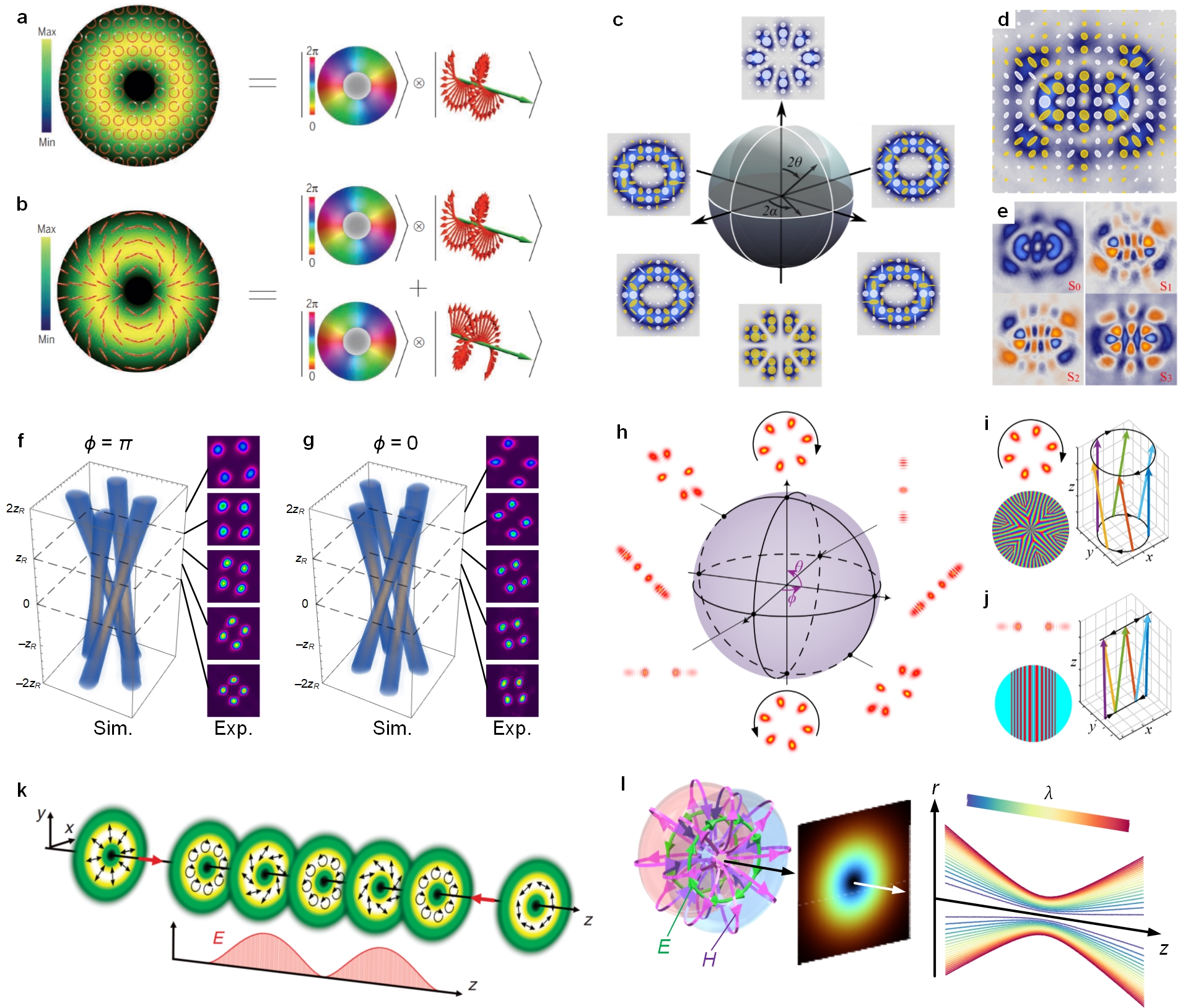}
\caption{\label{f3}{\bf\textsf{\mbox{Structured light analogs of typical quantum states.}}} \textbf{a,b}, A circularly polarized vector vortex beam is represented as the tensor product of a OAM state and SAM state (\textbf{a}), and a vector vortex beam is represented by a Bell state as the formation of OAM-SAM qubit entanglement (\textbf{b}) ~\cite{karimi2015classical}. \textbf{c-e}, The evolution of \rd{quantum-like} entangled elliptical mode can be mapped on a Poincar\'e sphere (c) and a Bell state of such mode is experimentally measured with polarization distributions (d) through the Stokes parameters (e)~\cite{Goldstein2011}. \textbf{f,g}, The multi-path ray-wave vortex beams with right- and left-handed twisted structures have the same formation of the quantum coherent state with coherent phase parameters of $\pi$ and 0~\cite{shen2018periodic}. \textbf{h-j}, The evolution of ray-wave nonseparable mode can be mapped on a Poincar\'e sphere (\textbf{h})~\cite{shen20202}, where the pole represents the circular twisted geometric mode carrying OAM (\textbf{i}), see the phase and spatial distributions inserted, and the equator the planer geometric mode (\textbf{j}). \textbf{k}, An \rd{state with an oscillatory degree of nonseparability} is constructed by a longitudinal-variant vector beam, that is superposed by two vortex beams with opposite propagation and orthogonal polarization~\cite{otte2018entanglement}. \textbf{l}, The flying doughnut pulse with focused single-cycle structure and isodiffracting spectrum fulfills the formation of space-spectrum \rd{quantum-like} entangled state~\cite{shen2021measures}.}
\end{figure*}

\subsection{Space-polarization nonseparable states}
In quantum mechanics, a qubit is the basic unit to encode information in bipartite systems, which is written as~\cite{nisbet2013photonic}:
\begin{align}
  \left| {{\psi }} \right\rangle = \alpha\left| 0 \right\rangle \left| 1 \right\rangle + \beta e^{i\varphi}\left| 1 \right\rangle \left| 0 \right\rangle 
\end{align}
where $\alpha^2+\beta^2=1$ and $\varphi$ is the phase delay between the two eigenstates. The maximally entangled states (leading to the strongest violations of Bell's inequality) in a bipartite qubit system are known as Bell states, and is the simplest case of entanglement of two particles~\cite{horodecki2009quantum}:
\begin{align}
  \left| {{\psi }^{\pm }} \right\rangle =\frac{\left| 0 \right\rangle \left| 1 \right\rangle \pm \left| 1 \right\rangle \left| 0 \right\rangle }{\sqrt{2}}\label{bell2}
\end{align}
In a \rd{nonseparable light} field, only local entanglement can be simulated using the intrinsic DoFs of a classical light field. For example, by selecting the OAM (spatial mode) and the SAM (polarization) DoF of light we obtain space-polarization nonseparable light modes with spatially variant polarization~\cite{karimi2015classical,aiello2015quantum,forbes2019classically}, as shown in Figs.~\ref{f3}\textbf{a,b}. Such 
qubit state can be expressed mathematically as:
\begin{equation}
    \left| {{\psi }} \right\rangle =\cos\theta|\ell\rangle|R\rangle + \sin\theta e^{i\varphi}|-\ell\rangle|L\rangle
\end{equation}
where $\theta$ is a parameter that allows a smooth transition between scalar ($\cos\theta=0$ or $\sin\theta=0$) and vector modes, ${\varphi}$ is an intermodal phase, $|\ell\rangle$ is an azimuthal index known as the topological charge  and $|R\rangle$, $|L\rangle$ are the states of right- and left-handed circular polarizations (RCP and LCP). Such vector beams can be mapped on the surface of a higher-order Poincar\'e sphere (HOPS)~\cite{milione2011higher}, revealing the quantum-like spin-to-orbital conversion (SOC). In this representation, \rd{there is a unique HOPS for each value of $\ell$, furthermore, it can be generalized to a hybrid-order Poincar\'e sphere by allowing modes with different topological charges $\ell_1$ and $\ell_2$~\cite{yi2015hybrid}. An even larger generalization can be done by using both, the azimuthal and radial indices that fully describe the transverse profile of LG modes.} The two polarization states can also be selected as any pair of elliptical orthogonal polarized states~\cite{devlin2017arbitrary}. {Moreover, the spatial DoF is not restricted to the set of LG modes but other sets of solutions to the paraxial wave equation can also be used. This is the case of the Ince-Gaussian~\cite{bandres2004ince,shen2019hybrid}, Mathieu-~\cite{Gutierrez-Vega2005} or parabolic-Gaussian\cite{Bandres2004} beams, which are solutions to the paraxial wave equation in elliptical and parabolic coordinates, respectively. In recent time, these and other families of modes have been employed to generate \rd{quantum-like} entangled beams with elliptical shapes, giving rise to new families of Bell-like vector novel types of  vector beams~\cite{yao2020classically,Rosales-Guzman2021,Hu2021,ZhaoBo2021}, see Figs.~\ref{f3}\textbf{c-e}. }

Space-polarization nonseparable states can be generated in various ways, for example using liquid crystal q-plates, which rely on the conversion of SAM to OAM ~\cite{marrucci2006optical,slussarenko2011tunable,marrucci2011spin,marrucci2013q}. A q-plate converts a circularly polarized Gaussian beam into a vortex beam of topological charge $\ell=\pm 2q$ and reverts its polarization direction, {$q$ is a parameter related to the number of cyclic rotations ($q$) around the center of the plate.} The sign of the topological charge is determined by the 
handedness of the input polarization state, see Fig.~\ref{f4}\textbf{a}. Hence, to generate a \rd{quantum-like} entangled mode, a linearly polarized beam is used as input, which is a superposition of RCP and LCP components. Another optical device recently proposed and with the capability to convert arbitrary elliptical polarization states to arbitrary output states of total angular momentum is the J-plate~\cite{mueller2017metasurface,devlin2017arbitrary}. The principle behind the J-plate relies on the use of subwavelength gratings with periods smaller that the wavelength of the input light, as shown in Fig.~\ref{f4}\textbf{b}. In recent years metamaterial approaches have been introduced that allow to realize vectorial states of increasing complexity, such as the metalens shown in Fig.~\ref{f4}\textbf{c}, which combine ondemand focusing to create multiple complex vector vortex states simultaneously~\cite{wang2019multichannel}. {In order to increase the compactness of the generating device as well as the purity of the generated modes, novel approaches have proposed the insertion of a metamaterial into the laser cavity~\cite{forbes2019structured},} (Fig.~\ref{f4}\textbf{g}) to directly generate high-purity nonseparable state with control over SOC even at extremely large topological charges~\cite{naidoo2016controlled,sroor2020high}.

Although compact, the metasurface as solid-state component is hard to reconfigure and does not allow modulation of light. As a recently emerged method, digital holography provides flexible, reconfigurable and programmable ways not only to generate arbitrary vector modes but also to access a higher dimensional polarization-OAM space. The first device that was used to this end was the liquid crystal Spatial Light Modulator (SLM), that even though it can only modulate linearly polarized states, it can tailor in an almost arbitrary way the spatial shape of light~\cite{schnars2015digital,forbes2016creation,rosales2017shape}. Hence, in order to generate \rd{quantum-like} entangled modes, the spatial DoF of each polarization component is manipulated independently after which they are recombined along a common propagation axis. To this end, several techniques have been proposed, based on one or two SLMs incuding the use of temporal sequences where a laser beam is bounced twice over a single SLM~\cite{maurer2007tailoring,otte2019polarization,otte2018spatial,moreno2012complete,rosales2017simultaneous}, as shown in Fig.~\ref{f4}\textbf{d}. Another device that is gaining popularity in recent years due to its many advantages, such as high refresh rates, polarization-independence, amongst others, is the Digital Micromirror Device (DMD)~\cite{ren2015tailoring,mitchell2016high,scholes2019structured,gong2014generation}, as shown in Fig.~\ref{f4}\textbf{e}. Notably, the polarization-independence property of DMDs allows the modulation of any polarization state, a property that has been widely exploited in recent times for the generation of \rd{quantum-like} entangled modes with arbitrary spatial shape and polarization distributions~\cite{selyem2019basis,rosales2020polarisation}. 

\subsection{Space-time nonseparable states}
In addition to qubit Bell states, high-dimensional Bell states have became increasingly topical~\cite{zhang2019arbitrary}, an $n$-dimensional case of which is given by:
\begin{equation}
    |\psi\rangle=\frac{1}{\sqrt{n}}\left(|0\rangle|0\rangle+|1\rangle|1\rangle+\cdots+|n-1\rangle|n-1\rangle\right)
\end{equation}
In quantum optics, there is a characteristic example of such Bell states, the OAM-entangled photon pair, $|\psi\rangle=\frac{1}{\sqrt{n}}\sum_{\ell=1}^n|\ell\rangle|-\ell\rangle$~\cite{erhard2018twisted}. But in the classical case, it is not possible to realize such states based on the polarization DoF, as it is limited in two-dimensions. The alternative route to this end is the selection of the two DoFs to form spatiotemporal structured light pulse, that is using space and time (or equivalently frequency or wavelength). Space-time non-separable pulses have a long history dating back 4 decades, where they were introduced in the context of non-diffracting waves~\cite{hernandez2008localized}. \rd{The method of termed ``electromagnetic directed-energy pulse trains'' was proposed as a well-endorsed way to solve general finite-energy space-time nonseparable pulses~\cite{ziolkowski1989localized}}, special cases of which were subsequently studied by Hellwarth and Nouchi, who found closed-form expressions that describe focused single-cycle pulse solutions to the homogeneous Maxwell’s equations. This family of pulses includes both linearly polarized pulses, termed ``focused pancakes''~\cite{feng1999spatiotemporal}, as well as pulses of toroidal symmetry, termed ``flying doughnuts''~\cite{hellwarth1996focused}, which were experimentally generated very recently~\cite{zdagkas2021observation}. The crucial characteristic induced by the space-time nonseparable pulse is the isodiffraction nature, whereby the spatial distribution of different spectral components in the transverse plane does not suffer distortion upon propagation.~\cite{feng1999iso,zdagkas2020space}, as demonstrated by the map of colored lines in Fig.~\ref{f3}\textbf{l}. Recently, a quantum-analogue method to characterize space-time (or -frequency) nonseparability was proposed, highlighting the link between isodiffracting pulses and high-dimensional Bell states. Particularly, spatiotemporal pulses can typically be described by a bipartite state with the space and time DoFs, but only specific type of space-time nonseparable pulses (e.g. isodiffracting pulses) can be the Bell state with maximal \rd{nonseparability}~\cite{shen2021measures}:
\begin{equation}
    |\psi\rangle=\frac{1}{\sqrt{n}}\sum_{i=1}^n|r_i\rangle|t_i\rangle
\end{equation}
or equivalently,
\begin{equation}
    |\psi\rangle=\frac{1}{\sqrt{n}}\sum_{i=1}^n|r_i\rangle|\omega_i\rangle
\end{equation}
where the spectral states $|\omega_i\rangle$ are (monochromatic) states of light of defined frequency $\omega_i$ and with defined radial position of peak intensity, and spatial states \rd{$|r_i\rangle$} are generally polychromatic states of light located at the position with defined radial ratio $r_i/r_\text{max}$ ($r_\text{max}$ is the radial position at which the total intensity of the light field reaches its maximum). For common pulses, the general state should be a general high-dimensional bipartite state, while the isodiffracting case corresponds to the maximally entangled state.

The exploration of the analogy between the space-time nonseparability and quantum entanglement would inject new vitality to the characterization of spatiotemporal structured pulses by using quantum-inspired methods such as state tomography and quantum measures. Space-time nonseparability can be employed for the construction of propagation-invariant spatiotemporal pulses~\cite{kagalwala2013bell,yessenov2019weaving,kondakci2019classical}, as well as the above discussed 
isodiffracting single-cycle pulses~\cite{shen2021measures}. Recently, the state tomography method is also used to measure more general kinds of classical wide-band structured beams or pulses~\cite{shen2021measures}. \rd{In addition, the study of space-time nonseparable states may not lbe imited in ultrashort pulses, which can also be accessed by the construction of propagation-dependent spatial mode, for instance, the polarization-variant polarization or OAM mode recently realized by metasurface design~\cite{dorrah2021metasurface,dorrah2021structuring}.}

\subsection{Ray-wave nonseparable states}
In addition to the Bell-like states, there are also many other intriguing bipartite quantum states available to be exploited. For example, the coherent state is a superposition of various Fock states, quantum state whose dynamics most closely resembles the oscillatory behavior of a classical harmonic oscillator~\cite{perelomov2012generalized}. In two-mode SU(2) symmetric system, the normalized SU(2) coherent state is given by~\cite{buvzek1989generalized}:
	\begin{equation}
	\left| \phi  \right\rangle =\frac{1}{{{2}^{{N}/{2}\;}}}\sum\limits_{K=0}^{N}{{{\left( \begin{matrix}
				N  \\
				K  \\
				\end{matrix} \right)}^{{1}/{2}\;}}{{\text{e}}^{\text{i}K\phi }}\left| K,N \right\rangle }
	\label{phi}
	\end{equation}
where $K$ and $N$ are number of bosons in the first and second modes of a Fock eigenstate $| K,N \rangle$ and $\phi$ is a phase parameter. As a salient property of coherent state, its wavefunction distribution is always coupled with the movement trajectory of the classical harmonic oscillator, which is termed quantum-classical correspondence~\cite{fox2000generalized}. The original proposal of coherent state involves the description of the photon stationary properties for distinguishing the quantum and classical field~\cite{perelomov2012generalized}, while in recent years it stands as a powerful tool to create geometric structured light. The SU(2) coherent state was originally used to characterize the statistic properties of photon flows in quantum optics. It can be used to describe structured light beams by replacing the Fock bases $|K, N\rangle$ by the classical light eigenmodes, either in freespace~\cite{chen2004wave,shen2018truncated,wan2020digitally}, waveguide~\cite{chen2012generation}, or microlaser systems~\cite{chen2002localization,chen2003vortex}. The resulting SU(2) geometric modes harness the striking property of \textit{ray-wave duality}, whereby the spatial wavepacket is always coupled with a geometric ray trajectory, akin to the quantum-classical correspondence of quantum coherent state. For example, the light modes in a microcavity, with billiard-like trajectory bouncing back-and-forth periodically within the boundary of cavity, can be represented by the coherent state field, and controlled by the parameters of coherent state~\cite{chen2002localization,chen2003vortex}. Such modes hold promise in the study of nonlinearity, chaos and non-Hermitian physics~\cite{bittner2018suppressing,cao2015dielectric,cao2019complex}. The coherent state field can also represent various exotic freespace geometric modes coupled with multi-path ray trajectory, moreover, the geometric ray trajectories can be tailored by control of a simple solid-state laser cavity~\cite{chen2013exploring,shen2018periodic}. {We can use the \rd{set of states}, $| K,N \rangle$, to represent a set} of frequency-degenerate freespace eigenmodes, e.g. HG, LG, and general HLG modes, {see Figs.~\ref{f3}\textbf{f,g} as an example of LG-based ray-wave geometric modes carrying OAM.} {Inspired by the OAM evolution of conventional HLG eigenmodes, the ray-wave geometric modes also feature an OAM evolution that can be mapped on a Poincar\'e-like sphere~\cite{shen20202}, as shown in Fig.~\ref{f3}\textbf{h}.} Due to the ray-wave duality structure, such geometric mode opens additional propagation-variant 3D spatial structures controlled by multiple parameters (see Figs.~\ref{f3}\textbf{i,j}) that the common beams do not have.

\subsection{\rd{Other two-DoFs nonseparable states}}
In contrast to the Bell-state vector beams, the higher-dimensional states of light provide access to a broader parameter space to manipulate customized patterns of light. A number of recent works have demonstrated control of higher dimensional beams based on the conventional vector beams. For example, the OAM state can be represented by the standing-wave mode, rather than the conventional traveling-wave LG mode. Thus it is formed by the superposition of two traveling-wave modes propagating in opposite directions, resulting into \rd{a beating of the degree of nonseparability}, i.e, a nonseparabl mode with an oscillatory degree of nonseparability along the longitudinal direction~\cite{otte2018entanglement}, as shown in Fig.~\ref{f3}\textbf{k}. This scheme was recently improved by using higher-order OAM modes for complete control of SOC dynamics~\cite{li2020spin,zhong2021}. The standing-wave eigenmodes can also be used to construct coherent state of light resulting into standing-wave ray-wave nonseparable structured light~\cite{wang2021unify}. The concept of vector beams were also transferred in spectral domain to create novel wavelength/frequency-polarization nonseparable structures~\cite{kopf2021spectral}. \rd{In addition to the above-mentioned DoFs, structured light can possess many other kinds of DoFs (as Fig.~\ref{f1} shows), and the combination of them can be used to construct more kinds of 2-DoF nonseparable states.} For instance along this line, pioneering works have demonsrated the generation of Bell states by combining linear momentum and OAM, i.e. momentum-OAM nonseparable state~\cite{pabon2020high}. \rd{In the field of nonlinear optics, it has been demonstrated that the use of path and polarization degrees of freedom allows nearly perfect optical switching between different OAM operations, such as, OAM addition and doubling \cite{Buono2018}.} 

In addition to the nonseparable states discussed above, vector beams have led to the generation of optical skyrmions. Skyrmions are special states of quasiparticles with a sophisticated space-vector topology, and in recent time, its optical counterparts were introduced from the quantum and material physics into optics, as a powerful tool to shape new topological states of light~\cite{tsesses2018optical,du2019deep,shen2021supertoroidal}. More recently, skyrmion topology also found its efficacy in tailoring reconfigurable patterns in vector beams by fully exploiting the salient particle-like topology of skyrmions~\cite{shen2021topological,shen2021generation,sugic2021particle}. Thus, more skyrmionic nonseparable states in structured light are expected in the future.

Finally, another promising path to follow is the construction of novel structured light fields based on other famous quantum states, \rd{such as the structured vortex modes of cat state~\cite{liu2019classical} and squeezed state~\cite{puentes2021generation}.} Therefore, it is highly probable that in the near future and with the exploitation of unexplored quantum-analogue methods, other kinds of bipartite or two-DoF higher-dimensional nonseparable light will be found.

\begin{figure*}
\centering
\includegraphics[width=0.95\linewidth]{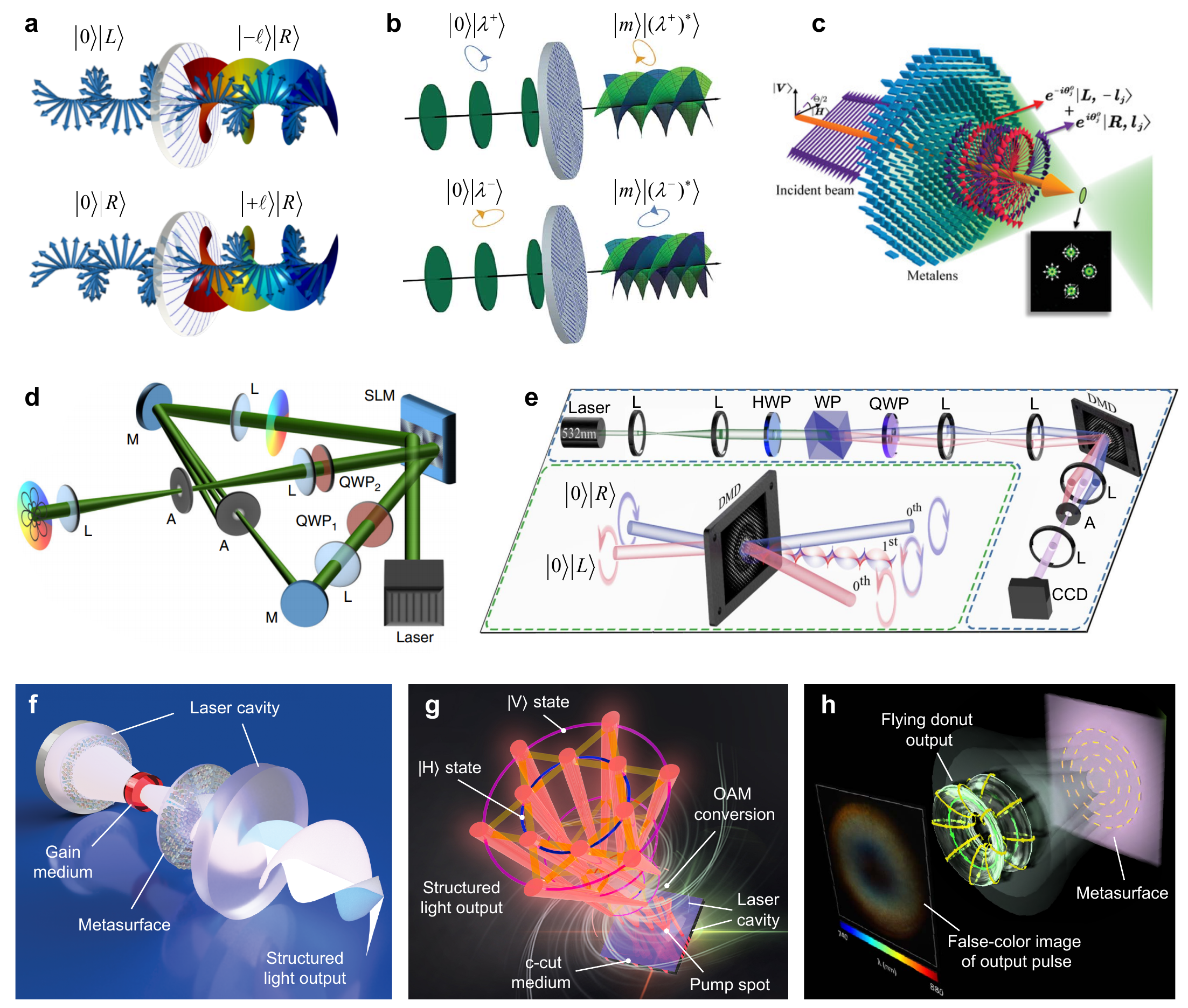}
\caption{\label{f4}{\bf\textsf{\mbox{Generation schemes of nonseparable states of light.}}} \textbf{a,b}, Schematic representation of the spin-orbit interaction performed by (\textbf{a}) a q-plate and (\textbf{b}) a J-plate~\cite{devlin2017arbitrary}. \textbf{c}, Multiple \rd{nonseparable vector modes} generated from a tailored metalens~\cite{wang2019multichannel}. \textbf{d}, Experimental setup to generate quantum-analogue modes using SLM~\cite{otte2019polarization}. \textbf{e}, Schematic representation of the generation of quantum-analogue modes using a digital micromirror device [104]. \textbf{f}, Generation of quantum-analogue modes directly from a laser with intracavity metasurface~\cite{forbes2019structured}. \textbf{g}, Generation of multipartite quantum-analogue states in more than two DoFs from an anisotropic microchip laser~\cite{shen2020structured}. \textbf{h}, Generation of quantum-analogue space-time nonseparable pulse (flying donut pulse) from a tailored metasurface with singular antenna array~\cite{quevedo2019roadmap}. L: lens; M: mirror; A: aperture; QWP: quarter-wave plate; HWP: half-wave plate; WP: Wollaston prism.}
\end{figure*}

\section{\rd{Multi-DoF nonseparable states}}
In previous sections, we outlined a roadmap to control \rd{quantum-like two-DoF} nonseparable states of structured light, by manipulating two intrinsic DoFs of light. Importantly, the use of higher-order modes, \rd{such as spatial modes with higher values of OAM ($\ell>2$), gives access to a larger dimension which allow to increase the information capacity per photon and the error threshold of optical communication systems. In regards to quantum entangled states, it has been demonstrated that the increase in dimension can translate into an increased robustness to noise, of potential applications to high-dimensional quantum key distribution protocols \cite{Ecker2019}.} Nonetheless, the mode order or OAM value that can be achieve experimentally has a technological limit. A way to overcome this limit is to exploit multiple DoFs of light to \rd{emulate multipartite entangled states}. \rd{For classical structured light, in recent decades, the nonseparable states were mostly discussed in vector beams, i.e. the space-polarization two-DoF nonseparable states, the construction of structured light with more-than-two DoFs is still elusive. Therefore, it is an emerging topic} to find more-than-two intrinsic DoFs of a structured light beam to construct on-demand multipartite nonseparable states. Based on the recent advances of structured light and optical manipulation, multiple new intrinsic DoFs of light have been proposed and controlled, that enables the construction of \rd{high-dimensional nonseparable states in multiple DoFs.} 

\subsection{Ray-wave-polarization nonseparable states}

The maximally entangled states that involve multiple subsystems are named Greenberger–Horne–Zeilinger (GHZ) states. \rd{In quantum entanglement, GHZ states require one DoF but in no-less-than-three DoFs.} For example the set of tripartite GHZ states that can be constructed as~\cite{pan1998greenberger}:
\begin{align}
  & \left| {{\Phi }^{\pm }} \right\rangle =\frac{\left| 0 \right\rangle \left| 0 \right\rangle \left| 0 \right\rangle \pm \left| 1 \right\rangle \left| 1 \right\rangle \left| 1 \right\rangle }{\sqrt{2}}\label{ghz1} \\ 
 & \left| \Psi _{1}^{\pm } \right\rangle =\frac{\left| 1 \right\rangle \left| 0 \right\rangle \left| 0 \right\rangle \pm \left| 0 \right\rangle \left| 1 \right\rangle \left| 1 \right\rangle }{\sqrt{2}} \\
  & \left| \Psi _{2}^{\pm } \right\rangle =\frac{\left| 0 \right\rangle \left| 1 \right\rangle \left| 0 \right\rangle \pm \left| 1 \right\rangle \left| 0 \right\rangle \left| 1 \right\rangle }{\sqrt{2}} \\ 
 & \left| \Psi _{3}^{\pm } \right\rangle =\frac{\left| 0 \right\rangle \left| 0 \right\rangle \left| 1 \right\rangle \pm \left| 1 \right\rangle \left| 1 \right\rangle \left| 0 \right\rangle }{\sqrt{2}}\label{ghz4}
\end{align}
\rd{All the 8 states are maximally entangled states. In general, a complete set of $m$-partite GHZ states include $2^m$ maximally entangled states. In order to construct the corresponding classical analogs of the GHZ states, we should find three different local DoFs to mimic the DoF in three subsystems. For instance,} the usage of mode indices of two orthogonal directions and polarization were applied to construct a tripartite \rd{GHZ-like} state~\cite{aiello2015quantum}. However, the two mode indices in a single beam cannot be tuned independently. By using the ray-wave vector vortex beam, as discussed in Section~3.3, it is possible to create more-than-two DoFs in a structured beam, such as trajectory shape, coherent-state phase, OAM, astigmatic degree and polarization~\cite{wang2021astigmatic}. Also, an experimental generation of tripartite nonseparable state with three of these new DoFs was reported, and, importantly, such modes can be generated directly from a laser cavity by operating an anisotropic microchip laser~\cite{shen2020structured}, as shown in Fig~\ref{f4}\textbf{g}. In this demonstration, the cavity lases on multiple ray-like trajectories or geometric paths, exploiting a ray-wave duality, with differing coherent state phases and independent tunable polarization. In addition, the polarization control was achieved via a precise off-axis pumping of an anisotropic crystal~\cite{guo2019anisotropic}. In this method, the creation of multiple DoFs and a single GHZ state is successful, but it is hard to fully control the high-dimensional nonseparable state. Recently, to overcome this bottleneck, it was reported that a complete set of tripartite GHZ states with all the eight maximally entangled states, as Eqs.~\ref{ghz1}-\ref{ghz4}, can be fully controlled in $2^3$-dimensional Hilbert space~\cite{shen2021creation}, by combing the spatial light modulation and laser control in multi-DoF ray-wave structured light. Moreover, this method is scalable to generate the multipartite GHZ state with even more controllable DoFs and higher dimensions.

\subsection{Space-time-polarization nonseparable states}

\rd{GHZ states are multipartite entangled states where the DoF is limited in 2-dimension. While, the general multi-DoF hybrid-entangled states can be constructed by different higher-dimensional DoFs, for example of a 3-DoF hyper-entangled states:
\begin{equation}
    \left| \psi  \right\rangle =\frac{1}{2\sqrt{n}}\left( \sum\limits_{i=0}^{n-1}{\left| i \right\rangle \left| i \right\rangle \left| 0 \right\rangle }+\sum\limits_{i=0}^{n-1}{\left|i \right\rangle \left| i \right\rangle \left| 1 \right\rangle } \right)
\end{equation}
where the first and second DoFs are $n$-dimensional, while the third DoF is 2-dimensional.}
An intuitive way to reach this goal is to find additional DoF in the structured light fulfilling high-dimensional Bell states. As discussed in Section~3.2, the isodiffracting single-cycle pulses can be expressed as high-dimensional Bell state with two DoFs (space and time). The class of isodiffractin pulses includes pancake-like pulses~\cite{feng1999iso,feng1999spatiotemporal} and doughnut-like pulses~\cite{hellwarth1996focused,zdagkas2020space}. The focused pancake pulse is linearly polarised, while the doughnut-like pulse presents a can perform complex structure of cylindrically polarized vector fields, including the azimuthal (for transverse electric modes)~\cite{raybould2017exciting} and radial polarized fields (for transverse magnetic mode)~\cite{zdagkas2019singularities} (the vector structure of transverse electric mode is depicted in Fig.~\ref{f3}\textbf{l}). Such cylindrical vector nature just performs as a platform to hold space-polarization nonseparable states~\cite{zhan2009cylindrical}, as the subsystem of the pulse. Thus the FD pulse has further properties to extend space-time-polarization nonseparable state:
\begin{equation}
    \left| \psi  \right\rangle =\frac{1}{2\sqrt{2}}\left( \sum\limits_{i}{\left| {{r}_{i}} \right\rangle \left| {{t}_{i}} \right\rangle \left| H \right\rangle }+\sum\limits_{i}{\left| {{r}_{i}} \right\rangle \left| {{t}_{i}} \right\rangle \left| V \right\rangle } \right)
\end{equation}
or equivalently,
\begin{equation}
\left| \psi  \right\rangle =\frac{1}{2\sqrt{2}}\left( \sum\limits_{i}{\left| {{r}_{i}} \right\rangle \left| {{\omega }_{i}} \right\rangle \left| H \right\rangle }+\sum\limits_{i}{\left| {{r}_{i}} \right\rangle \left| {{\omega }_{i}} \right\rangle \left| V \right\rangle } \right)
\end{equation}

Moreover, taking advantage of recent advances in the metasurfaces~\cite{quevedo2019roadmap}, the generation of such pulse was also proposed~\cite{papasimakis2018pulse}, {and observed experimentally in recent times}~\cite{zdagkas2021observation}, where the metamaterial converter can be constructed by a cylindrically symmetric array of low Q-factor dipole resonators oriented radially or azimuthally oriented electric dipoles and arranged in concentric rings, as shown in Fig.~\ref{f4}\textbf{h}. Such pulses have been shown to efficiently couple to toroidal multipoles and to non-radiating charge-current configurations (anapoles)~\cite{papasimakis2016electromagnetic,raybould2017exciting,raybould2016focused}. 
Recently, more new kinds of spatiotemporal vector pulses were designed~\cite{diouf2021space,guo2021structured,bliokh2021spatiotemporal,ChenWanChongZhan}, and the method for experimental characterization of ultrashort vector pulses was proposed~\cite{alonso2020complete,zdagkas2021spatio}, which can be used for fully measuring the distribution of the three DoFs, i.e. space, time, and polarization. Thus we can expect that the full control of tripartite nonseparable spatiotemporal pulse is just around the corner.

\subsection{Beyond \rd{three-DoF} nonseparable states}

\rd{The use of multiple degrees of freedom at the single photon level has been used to investigate the generation of entanglement between two noninteracting qubits that share a common environment, which have been prepared in independent individual states. Here, the two qubits where described by the transverse an polarization degrees of freedom, while the common enviroment was set by the photon path \cite{Hor-Myll2009}. In a similar way, achieving entanglement in high-dimensional quantum systems with one or more degrees of freedom is of great relevance to increase the information capacity or to enable new quantum information protocols \cite{Graffitti2020}.} Noteworthy, the classical analogs of multi-partite high-dimensional quantum states have already been suggested but controlling more than three DoF experimentally represents a real challenge. Recently, the structured light model to create up-to-five-partite GHZ states has been theoretically predicted~\cite{shen2021creation}, it requires the generation of more complex ray-wave geometric mode of higher-order coherent state so as to access two new DoF, center-OAM and sub-OAM, but has not been experimentally reported yet. A model of astigmatic hybrid SU(2) vector vortex beams was proposed to involve a new DoF, astigmatic degree, into ray-wave structured light~\cite{wang2021astigmatic}, but this DoF is still a candidate to be used to build nonseparable state. There were successful methods to control multipartite nonseparable state by designing a beam splitting system, where the increase of splitting paths reveals the creation of more parties~\cite{balthazar2016tripartite,spreeuw2001classical,balthazar2016conditional,qian2020quantification}. However, in this method, the DoFs would no longer be intrinsic to one paraxial beam and the control would become increasingly complicated and problematic. Therefore, it is still an open challenge, very much in its infancy, to find no-less-than four DoFs that are easy to control and intrinsic to a single paraxial structured beam.

Besides the quantum states selectively discussed above, there are still a number of states available to be explored for generating quantum-analogue structured light, such as the W state~\cite{sheng2012efficient} and Laughlin state~\cite{clark2020observation}, which all hold unique properties to be applied in classical light. Therefore, the methodology of constructing more kinds of quantum-like nonseparable state holds resource 
to realize multi-DoF higher-dimensional control of light for spurring new applications and extending the frontier of fundamental science.

\section{Applications of nonseparable states of light}

\noindent {Achieving applications of nonseparable states of light, fully or partially relying on the existing parallelism between these and quantum entangled states, is at the heart of future technologies.} Importantly, these are paving the road to applications that have demonstrated challenging with quantum states but can be easily implemented with quantum-analogue states. This is the case of various applications, where quantum inspired approaches started opening a new era of improvement and performance. Although, the unique properties of quantum-analogue modes of light have inspired many other applications (see for example Refs.~\cite{rosales2018review,shen2019optical}), here we will only focus on those that directly exploits the nonseparability aspect. 

\subsection{Beam quality measurement}
    
\begin{figure*}
\centering
\includegraphics[width=0.95\linewidth]{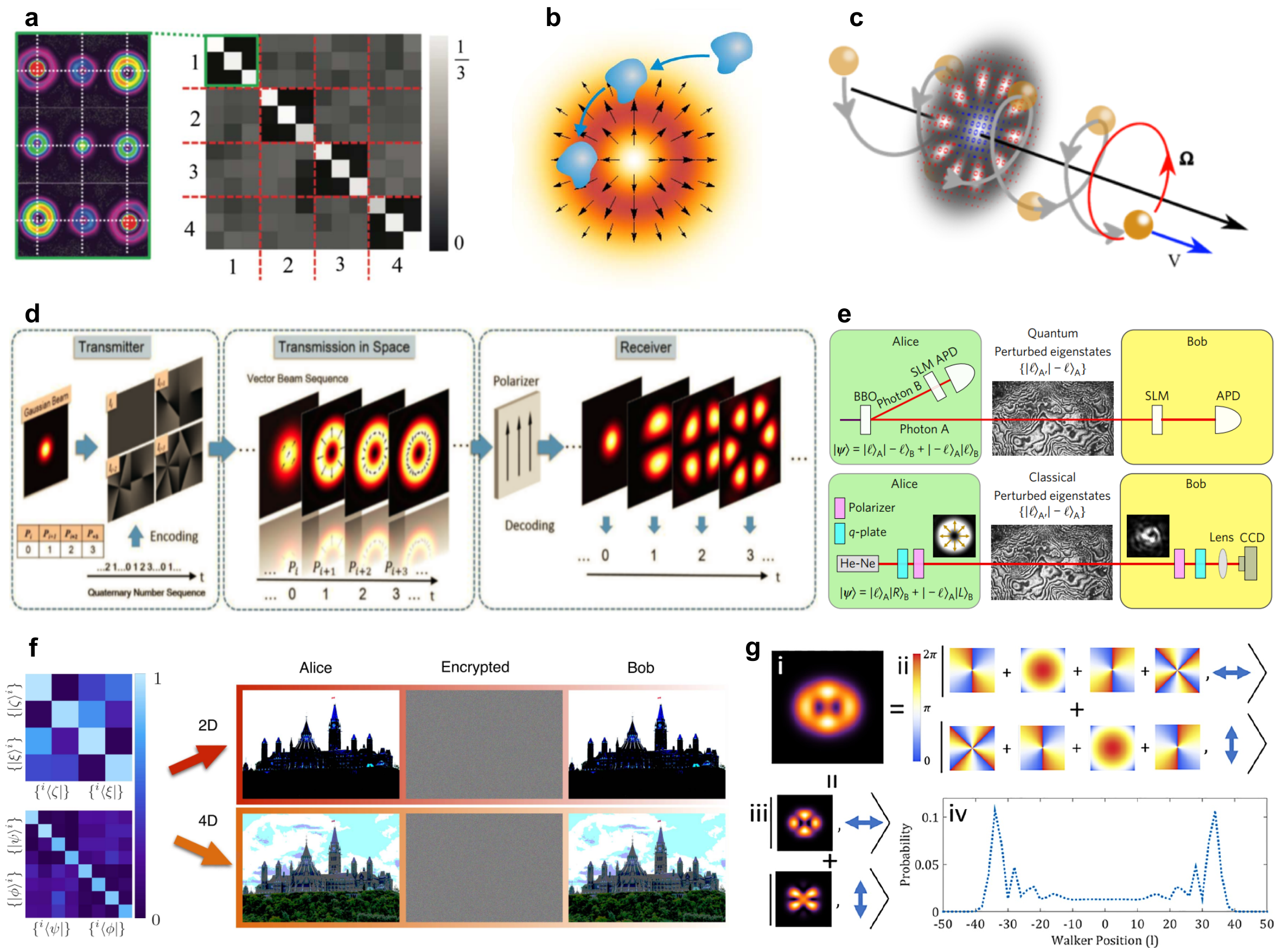}
\caption{\label{f5}{\bf\textsf{\mbox{Applications of nonseparable states of light.}}} \textbf{a}, State tomography measurement is implemented to measure the quality of classical structured light beams~\cite{toninelli2019concepts}. \textbf{b}, When an object traverses across a \rd{quantum-like nonseparable} beam, its spatial and polarization DoFs become a function of the instantaneous position of the object~\cite{berg2015classically}. \textbf{c}, The rotational and longitudinal speeds of an object moving in three dimensions can be measured simultaneously using a nonseparable light mode. \textbf{d}, Schematic of the transmission of a grey scale image using sixteen orthogonal nonseparable modes~\cite{zhao2015high}. \textbf{e}, Schematic of the transmission through a free-space turbulent channel of a quantum entangled photon and a \rd{quantum-like nonseparable} mode~\cite{ndagano2017characterizing}. \textbf{f}, Schematic of high-dimensional long-distance quantum cryptography for encrypting colorful images with structured light photon~\cite{sit2017high}. \textbf{g}, Schematic of the implementation of quantum walks using classical structured light: the intensity (i) and phase (ii) distributions of nonseparable superposition (iii) of OAM states and polarization for a 4-step walk, and the probability distribution (iv) of walker occupies positions after 50 steps of symmetrical walk~\cite{sephton2019versatile}.}
\end{figure*}

\rd{Since nonseparable light modes share the same mathematical formalism as quantum entangled states}, it is almost natural to exploit quantum mechanic tools to quantitatively characterize the corresponding properties of classical structured light. Importantly, Over the past century, an extended toolbox has been developed to quantify the quality of a quantum entangled state, including state tomography, density matrix, fidelity, linear entropy, concurrence, etc.~\cite{mqb,james2005measurement}. In the meantime, the measures for evaluating various properties of classical structured light is very limited, some original beam quality measures are only related to the scalar propagation property. While, the structured light has developed to harness multiple complex DoFs in itself, and new quantum-analogue measure methods are acquired to meet its development. For instance, the quantum state tomography measurement {can be applied to} vector vortex beams~\cite{toninelli2019concepts}, see Fig.~\ref{f5}\textbf{a}, to reconstruct the density matrix for characterizing the vector vortex beams and calculating the fidelity to reflect the quality to the ideal light state, akin to a measure of qubit~\cite{mclaren2015measuring}. We can also apply the concurrence or entanglement of formation, a measure of nonseparability of entangled state, to measure the quality of vector beams, as a new continual quantification from pure scalar beam to vector beam~\cite{ndagano2016beam,ZhaoBo2020,Selyem2019}. Moreover, the quantum Bell measurement method and Bell's inequality are also available to use for revealing the spin-to-orbital coupling in a classical beam~\cite{kagalwala2013bell,mclaren2015measuring}. Besides the pure state systems, the quantum mechanics of mixed state was also transferred to \rd{complex classical beams for exploring high-dimensional quantum-analogue system~\cite{zhou2021direct}}, which can also realize the counter-intuitive quantum teleportation effect in classical beams system~\cite{rafsanjani2015state,da2016spin}. \rd{In a similar way, the concept of nonseparability has been also used} for spatiotemporal quantum-analogue pulses, e.g. the Schmidt number or Schmidt rank was used to characterize the invariant propagation effect of a classical pulse~\cite{kondakci2019classical}. Since the quantum state tomography method was modified to \rd{high-dimensional state~\cite{zhou2021direct}}, we can apply analogous method to more complex structured light, e.g. the space-time nonseparable pulses, and many quantum concepts (fidelity, concurrence, linear entropy, etc.) were open to quantitatively characterize the spatiotemporal propagation dynamics~\cite{shen2021measures}.

\subsection{Metrology and sensing}

Optical metrology is a broad area of science and technology that encompasses all measuring techniques that use light as the main sensing tool, allowing for non-contact measurements at high speeds and with high precision. These can be used, for example, to measure the physical properties of solids and fluids, such as, their geometrical properties chemical composition, temperature, pressure, velocities, amongst other. In this regard, the nonseparability of quantum-analogue light beams provide with an alternative tool for the development of novel metrology techniques. This was demonstrated in a recent technique that exploits the nonseparability of such modes to perform real-time sensing of position and velocity of opaque objects moving in two dimensions~\cite{berg2015classically}. The key idea relies in the fact that when an object traverses across a \rd{nonseparable beam}, its space-polarization non-separability becomes a function of the instantaneous position of the object, see Fig.~\ref{f5}\textbf{b}. Importantly, the modulation of the spatial degree of freedom is correlated with the global polarization state of the \rd{nonseparable} mode for an extraction of the required information, directly from the changes in polarization. Along the same line, and also taking advantage of the nonseparability of such modes, a recent technique demonstrated the ability to measure simultaneously the longitudinal and angular speed of objects moving along spiral trajectories using a single interrogating beam~\cite{hu2019situ}. To this end, the moving target is illuminated with a \rd{nonseparable} mode constituted by a Gaussian and a vortex mode with orthogonal polarization, see Fig.~\ref{f5}\textbf{c}. The speed of the object is then determined from the light scattered back from the objects. Key to this is the fact that the information contained in the back-scattered light can be unambiguously separated upon detection. In this way, while the longitudinal speed is measured with the Gaussian component of the \rd{nonseparable} mode, the rotational one is determined with the vortex mode. 
The nonseparability of quantum-analogue light modes can be also used to determine the optical properties of objects. This was demonstrated in a recent technique that proposed a real-time Stokes polarimetry to measure the temperature of a temperature-dependent birefringent crystal~\cite{zhao2019real}. Stokes polarimetry enables the reconstruction of polarization through a minimum of four intensity measurements. In this way, a real-time reconstruction of polarization enables the possibility to track in over time the evolution of a given polarization-dependent property. In the experiment, a radially polarized beam was sent through a birefringent crystal while its temperature was increased causing a rotation of the polarization distribution. In this way, the angle of rotation of the state of polarization directly indicates the temperature of the crystal. In a similar work, the unique properties of \rd{nonseparable} modes were proposed as a two-DoF polarimetry scheme, a classical analogue of two-photon polarimetry. This with the aim of enhancing the performance of Mueller matrix metrology~\cite{toppel2014classical}. In its conventional form, Mueller matrix techniques allows to determine certain properties of an object by analyzing the polarization of light transmitted through or scattered from it. For this, the object must be sequentially illuminated with at least four different polarization states. In the new version implemented with quantum-analogue modes, the object is illuminated with a single light beam containing all polarization components, such as a radially polarized mode, in contrast to the conventional way that requires four beams. Afterwards, the light scattered or transmitted is analyzed simultaneously in both DoFs, polarization and spatial. Importantly, in a similar way to conventional Mueller matrix polarimetry, the polarization degree of freedom is used to prove the object while the spatial one is used to post-select the polarization state. This technique allows a real-time determination of the Mueller matrix allowing for fast tracking of the optical properties of the object. \rd{In addition to the space-polarization nonseparable vector modes, it is also an emerging direction to design other kinds of classical nonseparable state to enable novel metrology. For instance, the spectral vector mode as wavelength-polarization nonseparable state was designed and enabled a high-speed spectroscopic metrology~\cite{kopf2021spectral}.}

\subsection{Communications}

The realization that current communication systems will inevitably face a bandwidth capacity crunch, has risen an increasing interest in the development of alternative methods to encode and transmit information. Current techniques already exploit most of the physical properties of light, wavelength-division, time-division, quadrature amplitude modulation and polarization-division. Even though these techniques have significantly improved the transmission capacity of optical systems, most of them have reach their maximum capacity and will not be able to satisfy the exponentially-increasing global capacity demand. Importantly, mode division multiplexing provides a potential solution to this problem. Here information is encoded in the high-dimensional orthogonal space of the spatial modes of light, the shape of light. Due to their orthogonality, each mode can be used as a communication channel with the ability to carry independent data stream, increasing in this way the overall capacity in proportion to the number of modes. Despite all the advances along this research line (see for example~\cite{wang2016advances}), one of the major problems concerning both, free space and optical fibers communication is modal cross talk~\cite{trichili2020roadmap}. In free space, it is believed that \rd{space-polarization nonseparable} modes perform better in the presence of atmospheric turbulence than common scalar modes~\cite{cheng2009propagation,gu2009scintillation,gu2012reduction}, even though some studies contradict this believe~\cite{cox2016resilience}. The amount of research devoted to both, classical and quantum communication using \rd{quantum-like} modes is enormous~\cite{ndagano2017creation}. Pioneering work along this direction demonstrated, in a proof-of principle experiment, the transmission of information in free-space using four orthogonal \rd{space-polarisation nonseparable} modes, multiplexed and transmitted as one single light mode using mode division multiplexing. In this demonstration, each mode, generated with q-plates was able to transmit information at 20-Gbit/s~\cite{milione20154}. Similar techniques have demonstrated the transmission of \rd{such} light modes along optical fibers, which represents the core of our current communication systems~\cite{wang2017encoding,ndagano2015fiber}. An example of which is the transmission of data over 2 km using a fiber-based system that used vector modes generated from a photonic integrated device~\cite{liu2018direct,willner2018vector}. In addition to mode division multiplexing, \rd{nonseparable} modes can also be used as an  ``alphabet'' with which to encode and transmit information. In this approach, each vector mode of the \rd{multidimensional} basis is used as a symbol of the alphabet. In a first attempt to demonstrate this, four vector beams were encoded and successfully decoded with very low cross talk (lower 9\%) using a Mach–Zehnder interferometer~\cite{milione2015using}. An increase in the number of vector modes was achieved by Wang et al. who took advantage of the flexibility of SLMs to generate a basis containing 16 vector modes~\cite{zhao2015high}. With this, the authors were able to achieve hexadecimal encoding and used as an example the encoding and decoding of a $64 \times 64$ pixels black and white image, which was sent pixel by pixel. To this end, the grey value of each pixel is first converted to a hexadecimal number and then assigned to one of the 16 \rd{nonseparable modes}, see Fig.~\ref{f5}\textbf{d}. Similar experiments have demonstrated the realization of multi-ary encoding/decoding in a high-dimensional space formed by multiple possible states of hybrid vector beams~\cite{li2016high}. Recently, multi-DoF nonseparable ray-wave structured light was also exploited to improve the capacity, speed, and dimensionality in communication system~\cite{wan2021divergence}.  

\subsection{\rd{Information error corrections}}

The similarities between quantum and classical states of light can be exploited in the implementation of secure communications using the fundamental laws of quantum mechanics in what is known as Quantum Key Distribution (QKD). The secure transmission of information involves the implementation of a secure communication channel for the distribution of a cryptographic key. It is in this step where quantum mechanics provides an unconditional secure distribution of the cryptographic key by alerting from the presence of an intruder in the communication channel, based on fundamental laws of quantum mechanics, such as the non-cloning theorem and the Heisenberg’s uncertainty principle. Along this line, high-dimensional QKD protocols based on the spatial mode of light have become popular as they can potentially increase the transmissions distances as well as the information capacity~\cite{wang2016advances,ndagano2017creation}. Unfortunately, \rd{if we do not account for the perturbations in the quantum channel}, they can produce entanglement decay which, paradoxically, can only be repaired once the quantum link is working. A solution to this problem was proposed recently by taking full advantage of the similarity between quantum and spin orbit nonseparable states of light~\cite{ndagano2017characterizing}. In this proposal it was demonstrated that the quantum entanglement decay of a photon pair is identical to that of \rd{space-polarisation nonseparable multi-photon states}. This in the case where the channel acts on one of the two quantum entangled photons and similarly on one degree of freedom of the \rd{quantum-like nonseparable} mode, see Fig.~\ref{f5}\textbf{e}. In this way, a full characterization of a quantum channel can be performed using quantum-analogue light modes instead of single photons. This discovery provides the basis to perform precise quantum error correction in short and long-haul optical communication. Moreover, even though it was demonstrated for a free-space channel perturbed by atmospheric turbulence, it is expected to also work in optical fibers. \rd{Incidentally, in a recent article it was demonstrated that entanglement itself can also be used to measure the transmission matrix of the quantum channel, which in turn allows the recovery of quantum correlations that otherwise are lost \cite{Valencia2020}}.


\subsection{Computation and others}

Another advance of quantum effect is shown in the usage of computation, the quantum computation has proved its extremely higher speed and efficiency than the classical computation~\cite{divincenzo1995quantum}.  Since we can create various quantum-analogue states in classical light, it also inspires us to apply quantum computation algorithms and methods in classical optical computation system. A typical algorithm is the quantum walk~\cite{venegas2012quantum}. Quantum walks are motivated by the widespread use of classical random walks in the design of randomized algorithms for the usages of data retrieval and searching. The classical walks allow only one path in a retrieval process, while quantum walks promise to retrieve multiple paths simultaneously benefited by the quantum uncertainty nature, perfectly overcoming the classical drawback. Therefore, for some oracular problems, quantum walks provide an exponential speedup over any classical algorithm. Recently, complex structured light was used to implement a quantum walk and largely enhance its ability of data correction~\cite{preiss2015strongly,cardano2015quantum}. It was also proposed that the use of \rd{quantum-like modes allow to emulate the evolution of a quantum walk} in real time~\cite{goyal2013implementing}, which was also recently realized experimentally~\cite{sephton2019versatile}, see Fig.~\ref{f5}\textbf{g}. The classical quantum walk technique was also extended to higher-dimensional topological state for more powerful optical computations~\cite{d2020two}. \rd{Spin-orbit nonseparable states have been also used to illustrate some of the concepts of game theory in its quantum version, more precisely, to implement the quantum version of the prisoners dilemma~\cite{Pinheiro2013}. 
In addition, artificial intelligence and machine learning algorithms were recently used for identifying multi-DoF nonseparable states of light~\cite{giordani2020machine,wang2021deep}, which enable the further improvement of related applications. 
Finally, the use of nononseparable modes are taking relevance as an academic tool to simulate quantum key distribution protocols, see for example ~\cite{Souza2008,otte2020high} and references therein.} 

\rd{In short, in this review we highlighted many of the analogies that have been posted over the years between nonseparable light fields and quantum states but is very likely that we missed some other that are being proposed as we write this review. We also discussed a variety of applications, some of which are inspired  in the similarities between nonseparable fields and quantum states. Importantly, there are still some quantum effects which have not found their parallel with nonseparable states, some of which might never find it, these is the case, for example, of quantum imaging ~\cite{lemos2014quantum} or quantum storage~\cite{clausen2011quantum}.}
\section{Conclusions and perspectives}

In this review we outlined the many similarities that nonseparable states of classical light hold with quantum entangled states. To this end, we systematically constructed the theoretical framework of nonseparable states \cite{forbes2019classically,konrad2019quantum,shen2021creation}. Such \rd{entanglement-like states are represented mathematically} as a nonseparable superposition of the various DoFs of light, which can be classified akin to the classification of entangled states with different partite numbers and dimensionalities\cite{toninelli2019concepts}. This similarity performs as a bridge connecting the quantum and classical worlds and allows to apply the quantum tools to describe many of the properties of \rd{classical nonseparable states of light} \cite{mclaren2015measuring,Selyem2019,kagalwala2013bell}. Building on a large body of work on vector vortex modes, bipartite nonseparable states in polarisation and spatial mode, here we generalize the family of multipartite nonseparable state of light with diverse DoFs. With the recent technological advance, the family of nonseparable states of light is no longer limited to conventional vector beams, with an increasing numbers of intriguing higher-dimensional structured light modes (e.g. ray-wave structured light and space-time nonseparable pulse) enabling the simulation of an increasing number of higher-dimensional quantum states, which allows the transfer of more quantum tools to classical fields \cite{shen2021measures}. Therefore, we anticipate that in the following years, many other examples of quantum-analogue non-separable states of light will emerge.

Besides the techniques for the generation of quantum-analogue states of light, it is also important to develop techniques for the characterization of such states of light using quantum-inspired methodology. Pioneering work along this line, has demonstrated that many concepts, such as fidelity, concurrence, and linear entropy from quantum mechanics, which are used to measures of the degree of entanglement, can be applied also to \rd{space-polarisation nonseparable states} to measure the degree of coupling between the spatial and polarization DoFs in vector modes~\cite{ZhaoBo2020,ndagano2016beam}. Similarly, measures have been proposed to characterize space-time nonseparable structures, such as flying doughnuts~\cite{shen2021measures}. The techniques for the characterisation of quantum-analogue modes are still in its infancy and it is still challenging to adapt similar measurements to other DoFs. Nonetheless, the mathematical nature of entanglement does not distinguish between classical or quantum states, the only limitation is technological. Therefore it is very likely that technological advancements will provide the required tools to extend the concept of nonseparability to other novel DoFs, such as self-torque~\cite{rego2019generation}, astigmatic degree~\cite{wang2021astigmatic}, skyrmion number~\cite{shen2021topological}, supertoroidal order~\cite{shen2021supertoroidal}, in the near future. 

From the applications perspective, classically nonseparable states of light have pioneered many of them, some of which are directly inspired by their similarity with quantum entangled states.  Such applications, which have demonstrated challenging at the single photon level, are easier to develop with \rd{quantum-like classical nonseparable} states and will form the core of future technological advances. Nonetheless, the vast majority of applications revolve around the use of vector modes, \rd{nonseparable} in their spatial and polarisation DoFs, exploiting either their non-separability nature or their non-homogeneous transverse polarisation distribution. The development of applications of \rd{nonseparable states in} other DoFs has not been fully explored and will definitely set the basis for future technological advances.

Hence, even though \rd{many analogies} between the quantum and classical worlds have been established through the concept of nonseparability, only few DoFs have been fully exploited to explore such analogies from both, the fundamental and applications perspective. The zoology and methodology proposed in this article would surely guide new nonseparable states of light with more DoFs to be discovered. It is therefore highly probable that in the years to come quantum-analogue nonseparable states of light will merge as one of the main branches in modern optics.

\begin{acknowledgement}
  The authors thank Dr. Nikitas Papasimakis for useful discussions. CRG acknowledges support from the National Natural Science Foundation of China (61975047 and 11934013) and the High-Level Talents Project of Heilongjiang Province (Grant No. 2020GSP12).
\end{acknowledgement}

\begin{biographies}
  \authorbox{YS}{Yijie Shen}{is a senior research fellow in Optoelectronics Research Centre (ORC), University of Southampton, UK, also working as a Marie S.-Curie Research Fellow funded by MULTIPLY Marie S.-Curie Postdoctoral Fellowships. He received the Ph.D. degree in optical engineering at the Department of Precision Instrument, Tsinghua University, Beijing, China, in 2019. He received the B.S. degree in mechanical engineering and automation from South China University of Technology, Guangzhou, China, in 2015. During Mar. 2019 to Jun. 2019, he was invited as a visiting researcher in School of Physics: Structured Light Laboratory, University of the Witwatersrand (Wits University), Johannesburg, South Africa, and also invited as a visiting researcher in National Laser Centre, Council for Scientific and Industrial Research (CSIR), Pretoria, South Africa. His research interests include structured light, ultrafast nonlinear optics, quantum optics, metamaterial and nanophotonics.}
  
  \authorbox{CRG}{Carmelo Rosales-Guzm\'an}{Is a principal investigator at Centro
  	de Investigaciones en Optica (CIO, Mexico) since 2020 and an associate professor at the University of Science and Technology of Harbin (China) since 2018. He was a postdoctoral research fellow at the structured light laboratory of the University of the Witwatersrand (South Africa) from 2015 to 2018. He obtained his PhD degree in 2015 with the highest distinction {\it cum laude} from ICFO-The institute of photonics sciences (Barcelona, Spain). His main research interest involves the generation, characterisation and applications of structured light.  }  
\end{biographies}

\bibliography{refs}
\bibliographystyle{unsrt}
\end{document}